# Architecture of the Neurath Basic Model View Controller

**K. Yermashov, K. H. Siemsen, K. Wolke, R.A. Rasenack**
INK, FHOOW Emden and STRL, de Montfort University Leicester

**ABSTRACT.** The idea of the Neurath Basic Model View Controller (NBMVC) appeared during the discussion of the design of domain-specific modelling tools based on the Neurath Modelling Language [Yer06]. The NBMVC is the core of the modelling process within the modelling environment. It reduces complexity out of the design process by providing domain-specific interfaces between the developer and the model. These interfaces help to organise and manipulate the model. The organisation includes, for example, a layer with visual components to drop them in and filter them out. The control routines includes, for example, model transformations.

## 1. Introduction

The tools which use the NBMVC are supposed to have common basic functionalities (see figure 1), but differ by domain-specific features. These domain-specific tools are related to the system control software. They are called Input/Output devices Editor, Macro Editor and Task Editor. The common base for the three editors is, for example, the use of visual programming techniques and manipulating and transforming graph-like ASLT [WSY06] structures, which store, organise and annotate detailed model-related information. We develop a common framework, the NBMVC, which unites the common features of the three editors and which provides the environment to extend each editor with its domain-specific functionalities.

## 2. Common features

The three editors are domain-specific modelling tools which implement visual programming concepts. The Input/Output devices editor operates on the I/O elements to visually produce software Input and Output devices components. These components represent the correspondent hardware. Normally, Input and Output devices components are integrated together into a working application with the help of special controllers - we call them Macros. The Macro Editor operates with the I/O elements to visually produce Macro software components. The Input/Output devices Editor and the Macro Editor produce source code for the type or class entities. The concrete instances and how they work together according to the Macro schemes are set by the Task Editor.

All three editors have the following features in common:

- Use of Neurath visual programming. The model is composed from graphical symbols representing elements of the visual language. Graphical symbols can be sub-models themselves. Relations between elements are expressed with bindings.
- Use of ASLT data structures which are the underlying base for the model storage and code generation.

The difference between the three editors is dictated by their domain-specificity and includes the following points:

- Different model symbols for visualisation, layouts, groupings and filters.
- Different transformation rules to maintain the model.

## 3. Architecture

Figure 1 shows the architecture of the NBMVC. The base of the architecture is the Model-View-Controller (MVC) pattern [GHJ94]. The view part provides an interface to visually organise and manipulate a set of domain specific elements of the Neurath Modelling Language. The user works directly with the view. The controller part is the component which makes decisions and calculations, depending on events coming from the view part. The controller part directly modifies the data as shown in figure 1 by the data model part. The data model contains mainly an ASLT data structure, which holds properties related to the view part. As soon as the data model is changed it fires an event to the view part. Immediately the new model state is visually reflected by the view part.

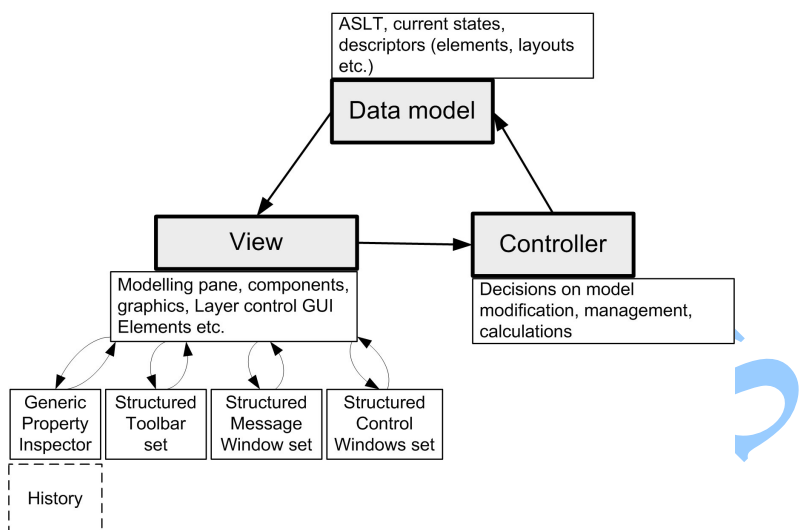

*Fig. 1:* NBMVC architecture

## 4. Event Processing

Next, we show some more elements the architecture consists of. We explain how events come to the view from outside (for example by user actions) and are sequentially processed. Figure 2 shows mode details of the NBMVC architecture. Arrows connect some elements and show the event movement according to the MVC pattern. The steps of event flow are numbered. The arrows 1…4 depict the event movement between elements before and during the data model modification. The arrows 5…7 depict the event movement between elements after the data model modification.

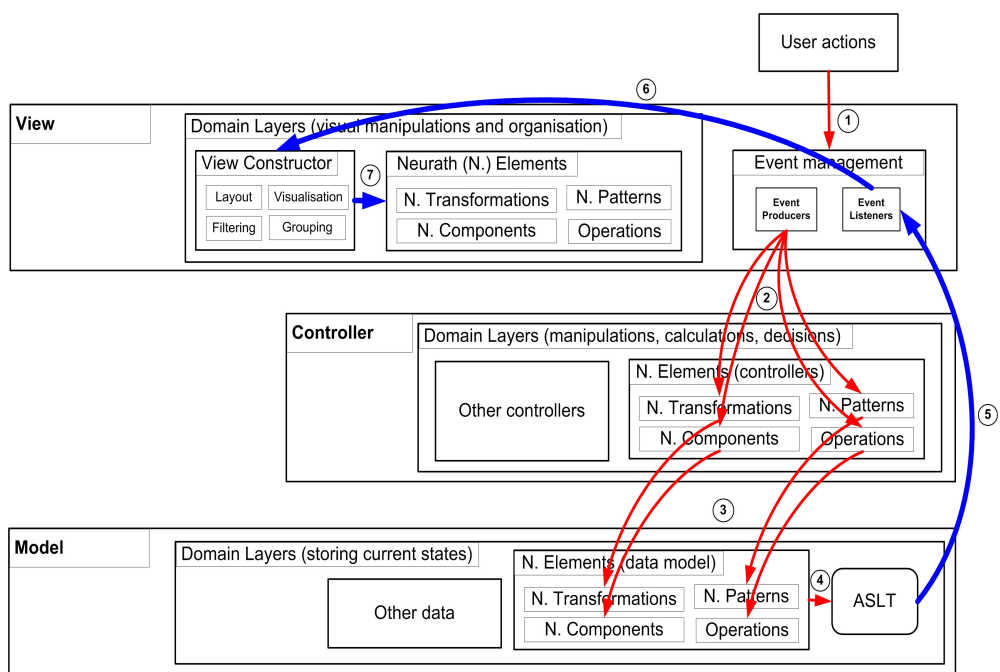

***Fig. 2:*** *Event processing within the NBMVC*

Initially, the MVC cycle within the tool based on NBMVC is born by one of the user actions. From one side, these actions can come from the modelling pane via, for example, mouse clicking or element dropping actions. From the other side, some events can come from another components, the modelling tool is consisted of, for example, from property inspectors. Anyway, the simple incoming events (for example, from the user or property inspector) are processed by the view in order to initiate more complex events which will correctly classify the change of the model.

Initialisation and fire of more complex events are shown as step two. The example of such complex events is INElementDroppedEvent, which is fired by the view when a component is dragged from the toolbar to the modelling pane. Normally, the typical event listener of all events coming from the view is the controller.

The controller's task is to analyse the incoming event with its event information, to make decisions and to initiate correspondent processes in order to modify the data model. Such model modification is depicted by the arrow with number three. A data model modification is, for example, a modification of the ASLT data structure. Step three is quite complex; it contains and uses a set of tools and sub-frameworks in order to map external domain-specific requirements. These requirements can specify, for example, which wizards are used and how they should be dynamically generated or what transformation processors should be dynamically instantiated. A lot of routines here have to do with the modification of the ASLT. We can specify here such libraries and components as domain-specific ASLT processors, atomic ASLT processors, and Meta Information processing tools (MIPTs).

Domain-specific ASLT processors do reflect some domain-specific transformation and are composed with a set of instructions which initialise atomic ASLT processors. Atomic ASLT processors use the MIPTs as a powerful and flexible framework to store and interact with the ASLT structure. Step four in figure 2 reflects the modification of the ASLT. Step four principally makes the newly generated or modified source code available.

The ASLT framework initiates the correspondent event instantly when the ASLT hierarchy or its values are changed. This event is shown with an arrow under number five. A typical listener for such events, according to the MVC pattern, is the view part.

The Neurath view part aims to represent the data model graphically. This task is managed by the View Constructor component. The arrow with number six depicts the event from one of the view event listeners to the View Constructor. The View Constructor produces the visual components within the view.

**Conclusion**

The NBMVC is a common general (not domain-specific) base for various domain-specific tools. It implements common visual programming features which include model storage, visualisation and transformation. These features can be used and extended by domain-specific tools. We specify three domain-specific editors based on the NBMVC model. The first is for defining the domain-specific input and output elements graphically. The second is for defining macro (control) elements visually. Input, output and macro elements each can be combined to produce extended elements with more functionality within their editors. The third editor is for combining inputs, macros and outputs by drag and drop to produce ready to use applications.

The NBMVC brings a basic reliability to the domain-specific tools and avoids the implementation of the same or similar routines many times.